\documentclass[a4paper,11pt]{article}
\usepackage{aaskaiid}
\usepackage{orcidlink}
\usepackage{aas_macros}
\usepackage{rotating}

\title{State-of-the-art Observation, Calibration, and Imaging Framework for Solar and Heliospheric Sciences with SKA}
\ShortTitle{Calibration and Imaging Framework for SH Science}

\author[1]{Divya Oberoi \orcidlink{0000-0002-4768-9058}} 
\ShortName{D. Oberoi et al.} 
\author[2,3]{Devojyoti Kansabanik \orcidlink{0000-0001-8801-9635}}
\author[1]{Soham Dey \orcidlink{0009-0006-3517-2031}}
\author[1]{Puja Majee \orcidlink{0000-0002-2711-2366}}
\author[1]{Surajit Mondal \orcidlink{0000-0002-2325-5298}}
\author[1]{Deepan Patra \orcidlink{0009-0003-8133-6621}}
\author[4]{Jingye Yan \orcidlink{0000-0002-7597-7663}}
\author[5]{Peijin Zhang \orcidlink{0000-0001-6855-5799}}
\author[6]{Pietro Zucca \orcidlink{0000-0002-6760-797X}}

\affiliation[1]{National Centre for Radio Astrophysics, Tata Institute of Fundamental Research, S. P. Pune University Campus, Ganeshkhind, Pune, India}
\emailAdd{div@ncra.tifr.res.in, div.oberoi@gmail.com}
\affiliation[2]{Johns Hopkins University Applied Physics Laboratory, 11001 Johns Hopkins Rd, Laurel, MD, USA}
\affiliation[3]{University Corporation for Atmospheric Research, 3090 Centre Green Dr., Boulder, CO, USA}
\affiliation[4]{State Key Laboratory of Solar Activity and Space Weather, National Space Science Center, Chinese Academy of Sciences, Beijing 100190, China}
\affiliation[5]{Center for Solar-Terrestrial Research, New Jersey Institute of Technology, Newark, NJ 07102, USA}
\affiliation[6]{ASTRON,  The Netherlands Institute for Radio Astronomy, The Netherlands}

\abstract{
The Sun is a surprisingly difficult radio source to observe and image, even with the SKA. It is multiple orders brighter than the typical radio sources, which sensitive radio telescopes like SKA are optimized for. So, configuring the signal chain to enable solar observations while maintaining linearity is the very first non-standard requirement to be met. Next, solar radio emission spans an impressive range along every single phase-space parameter that can be used to describe it -- time scales from solar cycles to millisecond; spectral scales from smooth thermal emission to $\sim$100 kHz coherent emission; brightness temperatures from $10^4$ K for gyrosynchrotron emissions to $10^{13}$ K for bright type-III bursts; fractional polarizations from less than 1\% to nearly 100\%; and angular scales extending beyond a degree. Capturing the dynamics in solar radio emission in their full glory requires, on the one hand, that all the data that goes into making an image be acquired over very short temporal and spectral spans and, on the other, also imposes requirements for very high imaging dynamic range with high polarization purity. Extracting the information at the requisite temporal and spectral scales from SKA data will require a spectropolarimetric snapshot capability with high dynamic range and fidelity. Additionally, some of the most interesting insights into solar physics and space weather come from studying solar activity, which remains inherently unpredictable. This chapter discusses the various considerations that need to be addressed to help realize the promise of solar and heliospheric science from SKA.

}

\begin{document}
\maketitle

\section{Introduction}
High fidelity observations of the Sun place some unusual demands on the radio interferometric calibration and imaging framework usually adopted for the modern state-of-the-art interferometers. Although these instruments are designed to be versatile in their capabilities and address a broad range of science objectives they can pursue, the extreme observational requirements of solar science tend to present {\em corner cases}. This leads to situations where, even though intrinsically the instrument hardware is quite capable of capturing the information well matched to the needs of solar science, the limitations posed by the downstream signal chain and the functionality for which the calibration and imaging pipelines are optimized end up making the final data products less than optimal, or even unusable, for solar science.
Pursuing solar and heliospheric science with these instruments, hence, requires considerable additional effort. 
The natural emphasis on first making accessible a large part of the less challenging observations phase space before turning to the corner cases, especially given the constraints due to typically lean debugging and commissioning teams and the comparatively small size of the solar radio community, lead to the solar observing modes being among the last ones to be commissioned on most advanced interferometers (e.g. ALMA, MeerKAT and ASKAP).
It is important to acknowledge that the issues associated with solar observing are intrinsically harder at higher frequencies and a lot more tractable for the wide field of view aperture arrays at lower frequencies.
The SKA-Low precursor, the Murchison Widefield Array (MWA), included solar and heliospheric science as among its key science objectives from its inception, yielding rich science dividends \citep{Oberoi2011, Oberoi2023}.
Despite all of the difficulties associated with enabling solar and heliospheric science with the modern state-of-the-art interferometers, like the SKA telescopes, these instruments are intrinsically a far better match to the needs of solar imaging, 
both for the quiet \citep{Mondal.1.2026.SKA} as well as active regions and emissions \citep{Kumari.1.2026.SKA,Morosan.1.2026.SKA, Patra.1.2026.SKA}, as compared to the solar dedicated radio observatories (e.g., Nan\c{c}ay Radio Heliograph, Gauribidanur Radio Heliograph, etc.) -- completely unsurprising given the far limited resources available to the latter.
To enable solar and heliospheric science from the very start of the SKA's science operations, it is of crucial importance to carefully identify and address the many different aspects needed for enabling solar and heliospheric observations with the SKA.

Here we collate the most important of these aspects along with brief descriptions of how they have been addressed at the various SKA precursors and pathfinders and similar new-generation radio interferometers which have been used for solar and heliospheric science.
In addition, we also list some other aspects specific to solar and heliospheric observations which high-fidelity and high-sensitivity imaging at very high spectro-temporal and spatial resolution will need to contend with. This paper is organized as follows -- 

\section{Adapting the analog signal chain for solar observations}
\label{sec:analog-signal-chain}
\begin{figure}[!htbp]
    \centering
    \includegraphics[trim={0cm 2.5cm 0cm 2.5cm},clip,width=0.7\linewidth]{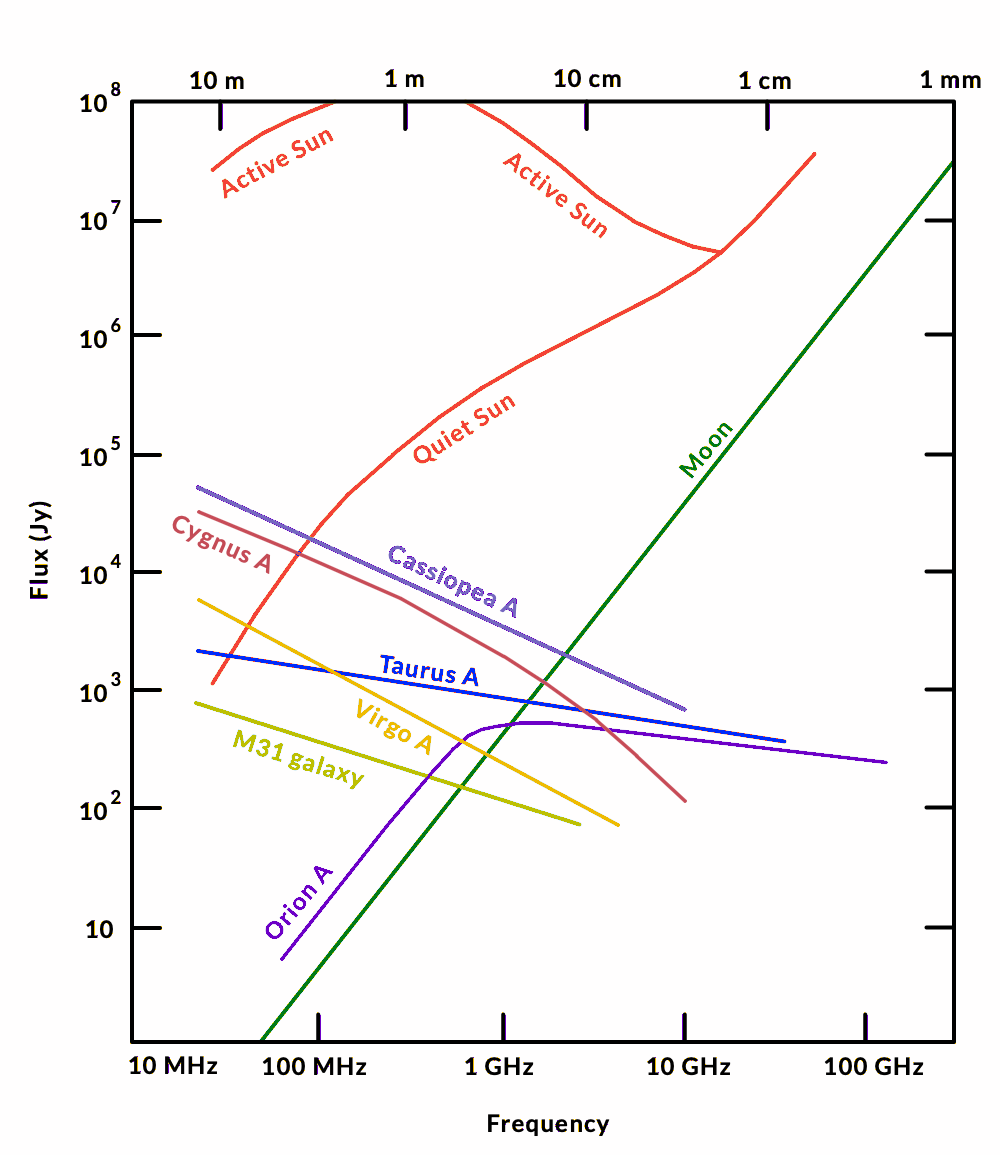}
    \caption{Radio spectra of bright radio sources in the sky, and that of the quiet and the active Sun. (Adapted from \citet{kraus_radio_astronomy_1986}).}
    \label{fig:radio_spectra}
\end{figure}

The state-of-the-art interferometers are optimized for observations of very faint sources, with the objective of routinely getting down to sub-$\mu Jy$ levels. 
The Sun, on the other hand, simply because of its close proximity, is by far the radio source with the highest flux density in practically the entire SKAO frequency range.
The flux density of even the quiet Sun lies in the range from $\sim$1 to $\sim$1000 SFU (1 SFU = $10^4$ Jy) from the low end of the SKA-Low to the high end of SKA-Mid, and increases by multiple orders of magnitude during periods of activity. 
Building an analog signal chain with a linear response regime large enough to span the range from typical astronomical sources to the very strong and highly variable solar signal becomes increasingly challenging.
The fact that the quiet solar flux density increases with frequency, while that of typical astronomical calibrator sources emitting synchrotron emission decreases with frequency, as shown in Figure \ref{fig:radio_spectra}, makes this even more challenging at higher frequencies.
This section describes how the analog part of the signal chain is modified for solar observations with some examples from SKA-Low and SKA-Mid precursors and pathfinders, and recommendations for SKA telescopes. 

\subsection{SKA-Low precursors and pathfinders -- MWA and LOFAR} 
The problem of adapting the analog signal chain for solar observation is admittedly less severe at low frequencies. 
Low-frequency instruments, like the Murchison Widefield Array \citep[MWA;][]{lonsdale2009,tingay2013}, LOw Frequency ARray \citep[LOFAR;][]{lofar2013}, and also SKA-Low, use aperture arrays which naturally lead to wide fields-of-view (FoV). 
Additionally, the Galactic background emission tends to get brighter as one progresses to lower frequencies \citep{Rogers2008}. 
It is convenient to discuss this quantitatively in terms of the effective increase in the system temperature, $T_\mathrm{sys}$, due to the presence of the Sun in the FoV.
Though the Sun has a very high brightness temperature \footnote{$T_\odot$, the disk averaged brightness temperature of the Sun approaches $10^6\ K$  for the quiet Sun at meterwavelengths, which can grow as high as $10^{13}\ K$ for compact narrowband active emissions}, its angular size is very small when compared to the primary beam size of these instruments.
The FoV of MWA tiles is $610\ deg^2$ at $150\ MHz$ \citep{tingay2013}, while the Sun's angular size at the same frequency is $\sim 0.3\ deg^2$, leading to a beam dilution by a factor of about 2000. 
Presence of the Sun in the FoV only adds a few hundred $K$ to the $T_\mathrm{sys}$, and is comparable to the contribution of the Galactic background to $T_\mathrm{sys}$ \citep{oberoi2017}. 
At MWA, an additional attenuation of $10\ dB$ is introduced in the analog signal path for solar observations, motivated by the desire to maintain some headroom in the 5-bit analog-to-digital converters (ADCs) \citep{tingay2013} for the potentially large additional power contributed by solar radio bursts. 
LOFAR uses ADCs with higher bit-depth (8 or 16 bit, depending on the number of subbands in use), to be able to deal with the presence of much stronger radio frequency interference (RFI) expected in its environment.
The LOFAR analog signal has sufficient headroom to accommodate even bright solar bursts, and no additional attenuation needs to be introduced for solar observations.

For the SKA-Low, the beam size will be governed by the effective station diameter of $38\ m$\footnote{\href{https://www.dropbox.com/scl/fi/4aiatav4gqxh3hwom7901/SKA-TEL-SKO-0001075-02_DesignBaselineDescription.pdf?rlkey=8qme7b9spoo21s0hlda28li0t&e=1&st=vjy1oa2k&dl=0}{SKA Baseline Design Description}}.
A simple scaling suggests that the solid angle of the SKA-Low primary beam will be $\sim60$ times smaller than the $\sim 5\ m \times 5\ m$ MWA tiles, leading to a reduction in beam dilution of $T_{\odot}$ by a similar fraction.
Also, SKA-Low observing range extends to comparatively higher frequencies as compared to the MWA or LOFAR, where the solar flux density increases, and that of the Galactic background falls.
The SKA-Low will hence need to provision for introducing significant attenuation in the analog signal path to enable solar observations. 
The amount of attenuation required will be a function of frequency and will depend upon the details of the analog signal chain, the number of bits used at the ADC, and the downstream signal processing.
We note that SKA-Low can also construct subarrays using sub-stations \citep{SKAO-Low-SubStation-Templates}. 
Limiting the footprint of the dipoles in a station will yield a larger FoV and proportionally larger beam dilution. Though it will come at the cost of reduced sensitivity, it might still be an acceptable or even a desirable trade-off for solar and heliospheric observations as discussed in Section \ref{subsec:triggered-obs} and \ref{subsec:adaptive-fov}, respectively.

\subsection{SKA-Mid precursors and pathfinders -- MeerKAT and ASKAP} 
\label{subsec:analog-MeerKAT}
For instruments at high frequencies which use filled dishes, like MeerKAT, Australian Square Kilometre Array Pathfinder \citep[ASKAP;][]{askap2021} and the SKA-Mid, the problem is more challenging -- the Sun not only fills a substantial part of the primary beam, its flux density is also much larger than at lower frequencies.
Preparing for solar observations has to start with ensuring that the roughness of the dish reflector surface is sufficiently large, so that it can diffuse the visible and infrared sunlight being focused on to the receiver and prevent any heat damage. For example, at the ASKAP, there was an incident where deposition of dew on the dish surface made it smooth enough to concentrate sunlight at the primary focus and damaged the phased-array feeds. MeerKAT dishes have been designed to have sufficient surface roughness as well as the receiver at the secondary reflector. Hence, it can be pointed directly at the Sun \citep{kansabanik2025solar}. We expect not to face this issue for the SKA-Mid dishes, as they are similar to those at MeerKAT.  

The low-noise-amplifiers (LNAs) - the very first component in the signal chain - themselves usually have a sufficiently large dynamic range to accommodate even the active solar emissions in the linear regime of operation. This has been shown to be the case for MeerKAT \citep{kansabanik2025solar} and is also expected to be true for the SKA-Mid.
The dynamic range of the downstream signal chain is much more limited and, hence, requires careful and non-standard configuration to accommodate the much stronger solar signal. In fact, the early solar observing efforts with MeerKAT attempted to achieve the required attenuation by observing the Sun in sidelobes of the primary beam \citep{Kansabanik_2024_meerkat}. 
A more recent attempt, pursued with the objective of establishing a solar observing mode with MeerKAT, made use of the Radio Frequency Conditioning Unit (RFCU) present in the signal chain just prior to digitisation. RFCU provides the ability to attenuate the analog signal by up to 63~$dB$ in steps of 1~$dB$, to observe the Sun in the main lobe of MeerKAT primary beam. In the MeerKAT UHF (580–1015 MHz) and L-bands (900–1670 MHz), where solar observations have been validated thus far, introducing an additional attenuation of $\sim$~32~$dB$ and $\sim$~35~$dB$ respectively was found to be sufficient for quiet Sun as well as active solar conditions \citep{kansabanik2025solar}.

Solar observations with the SKA-Mid will require the ability to include additional frequency-dependent attenuation in the analog part of the signal chain. This additional attenuation is expected to lie in the range from $\sim$~32~$dB$ in the UHF band, growing to about $\sim$~40~$dB$ by 15~$GHz$, and the ability to adjust attenuations in steps of 1--2~$dB$ will be adequate.

\section{Characterisation of additional attenuators for solar flux density calibration}
\label{sec:flux-calib}
Radio interferometers rely on observations of a handful of flux density calibrators to establish the flux density scale.
The fact that the analog part of the signal chain needs to be configured differently for calibrator and solar observations implies that the usual approach of determining the antenna gains using calibrator observations and transferring them to the target source is no longer feasible. As expected, the situation gets progressively more challenging at higher frequencies, with the flux density of the synchrotron-powered calibrator sources dropping and the thermal solar flux density rising, as shown in Figure \ref{fig:radio_spectra}.
In principle, all one needs to establish the flux density scale is a precise {\em a priori} estimate of the additional attenuation introduced in the signal path.
Once this additional attenuation is compensated for, the rest of the flux density calibration process can proceed as usual. 
In practice, however, sufficiently precise characterisation of attenuators is rarely available. Even though laboratory measurements of attenuators might be available, their behaviour on the telescope might get modified or evolve with time across the lifetime of the instrument. This section summarizes the techniques that have been developed to achieve reliable solar flux density calibration for both SKA-Low and -Mid precursors and pathfinders, along with recommendations for the SKAO telescopes. 

\begin{figure}[!htbp]
    \centering
    \includegraphics[trim={0cm 0cm 8.9cm 0.32cm},clip,width=0.42\linewidth]{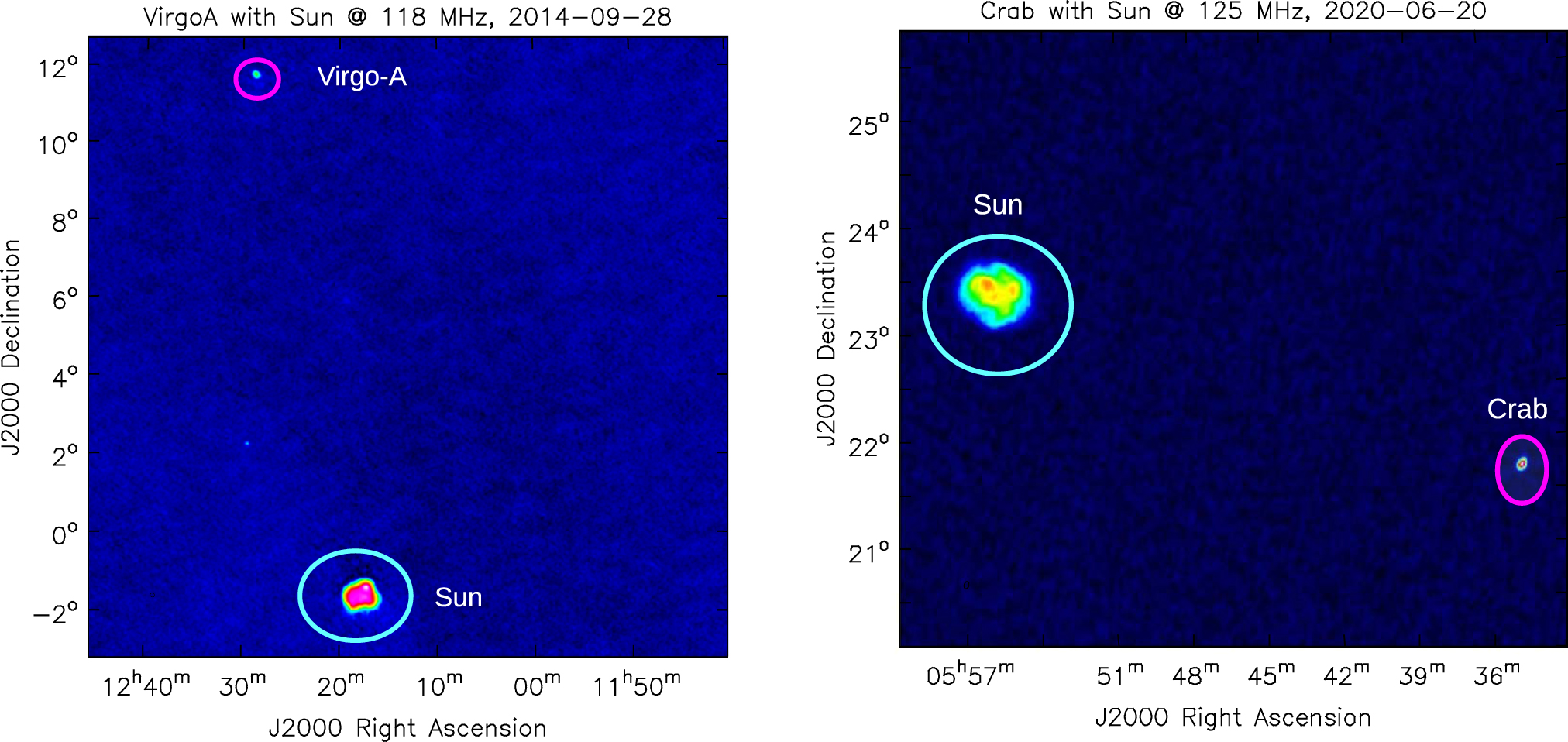}\includegraphics[width=0.4\linewidth]{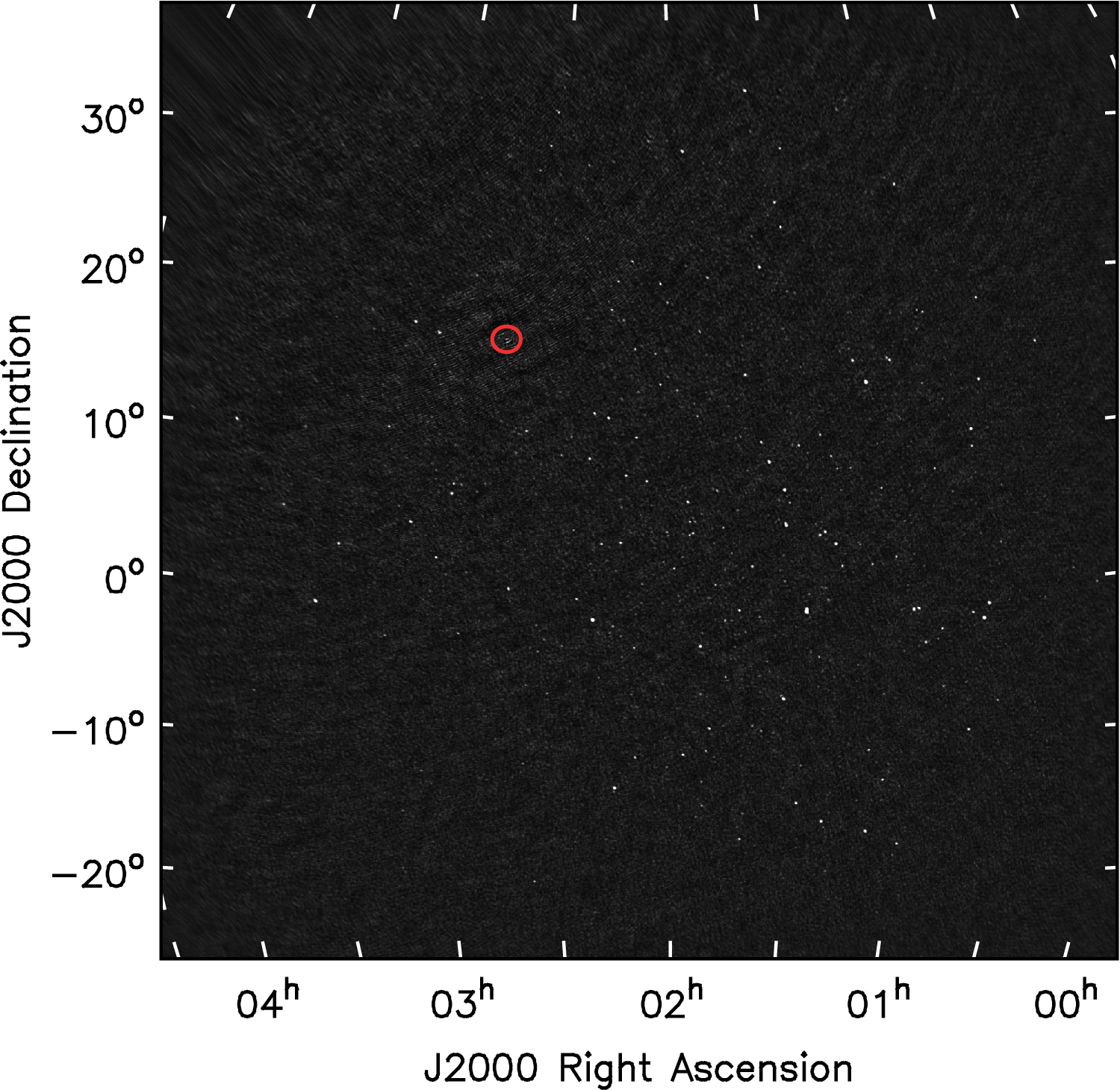}
    \caption{Left panel: A MWA map at 118 MHz showing Virgo-A (A-team source) and the Sun in the same FoV. Right panel: Image centered at 80 MHz, $\sim60^\circ$ on each side, obtained after subtraction of solar emission from the data. The red circle with radius $2\ R_\odot$ is the region where the Sun was present (Reproduced from \citet{Kansabanik2022}).}
    \label{fig:source_with_sun}
\end{figure}

\subsection{SKA-Low precursors and pathfinders -- MWA and LOFAR} 
At the MWA, the first approach used for flux density calibration was essentially an extension of the approach of using sky models for calibrating individual antennas to a two-element interferometer \citep{oberoi2017}. 
It relied on the availability of a good model for the antenna primary beam, the sky, and a sufficient number of baselines short enough not to completely resolve the large angular scale emission from the Galactic background. This approach worked well for the MWA Phase I.
The MWA phase II \citep{Wayth2018} did not have a sufficient number of baselines short enough for this approach to work reliably, particularly at the higher part of the observing band, and alternative strategies needed to be developed. 
One approach was to use observations of the so-called {\em A-team} sources with and without the attenuation used for solar observations to effectively characterize the attenuator response. Alternative methods have relied either on the chance presence of bright background sources within the same field of view as the Sun (Figure \ref{fig:source_with_sun}, left panel) or on achieving sufficiently high Stokes I imaging dynamic range to enable detection of multiple background sources in the solar field of view \citep{Kansabanik2022} (Figure \ref{fig:source_with_sun}, right panel). In both cases, the solar flux density is estimated by referencing the catalogued flux densities of these background sources. For LOFAR, as no additional attenuation is needed for solar observations, the process of flux density calibration also does not require any modification. The approach used for the MWA is expected to work similarly for SKA-Low solar observations.

\subsection{SKA-Mid precursor -- MeerKAT}\label{subsec:meerkat_fluxcal}
\begin{figure}[!htbp]
    \centering
    \includegraphics[trim={2cm -3.3cm 0cm 0cm},clip,width=0.5\linewidth]{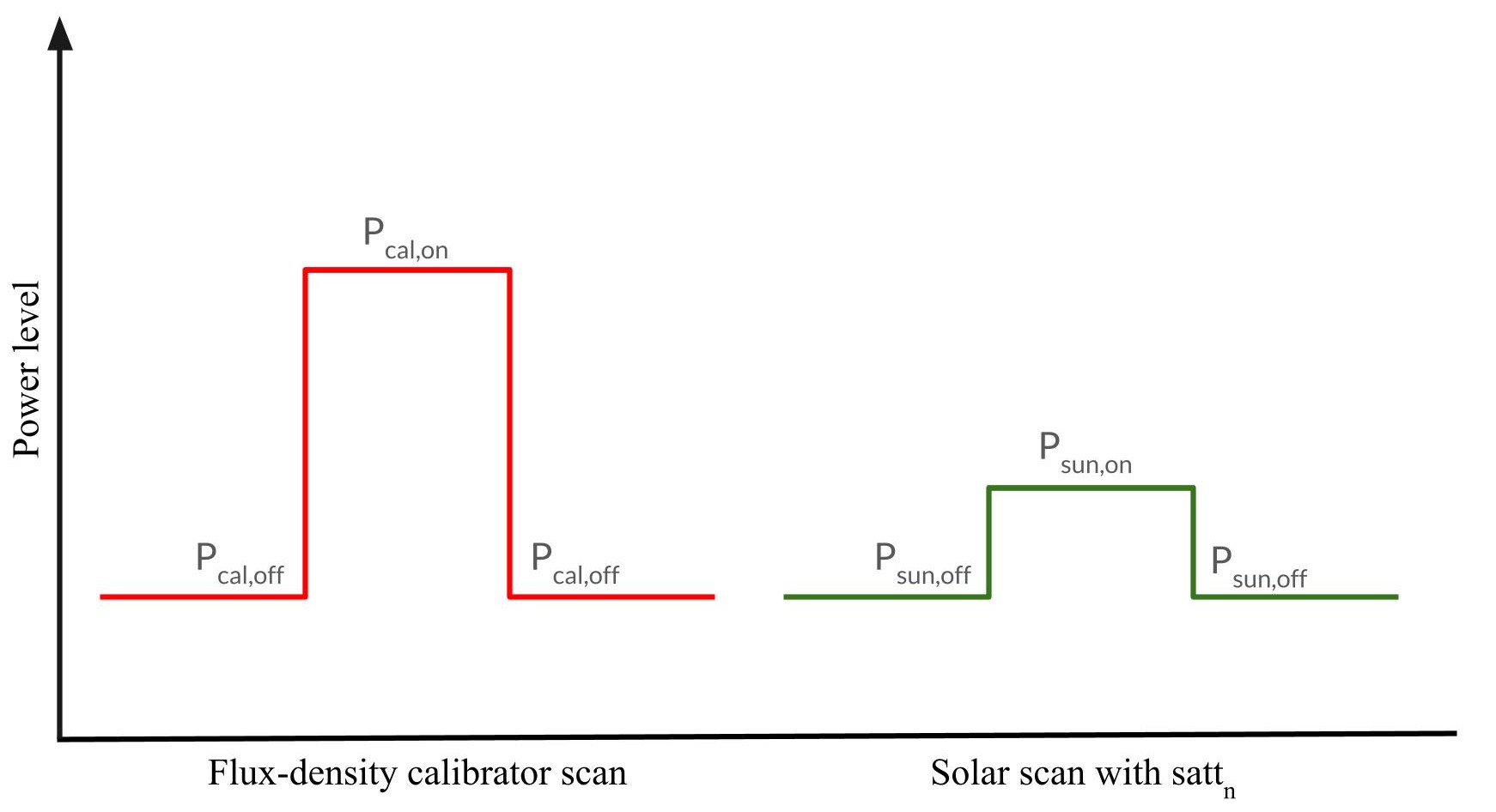}
    \includegraphics[width=0.4\linewidth]{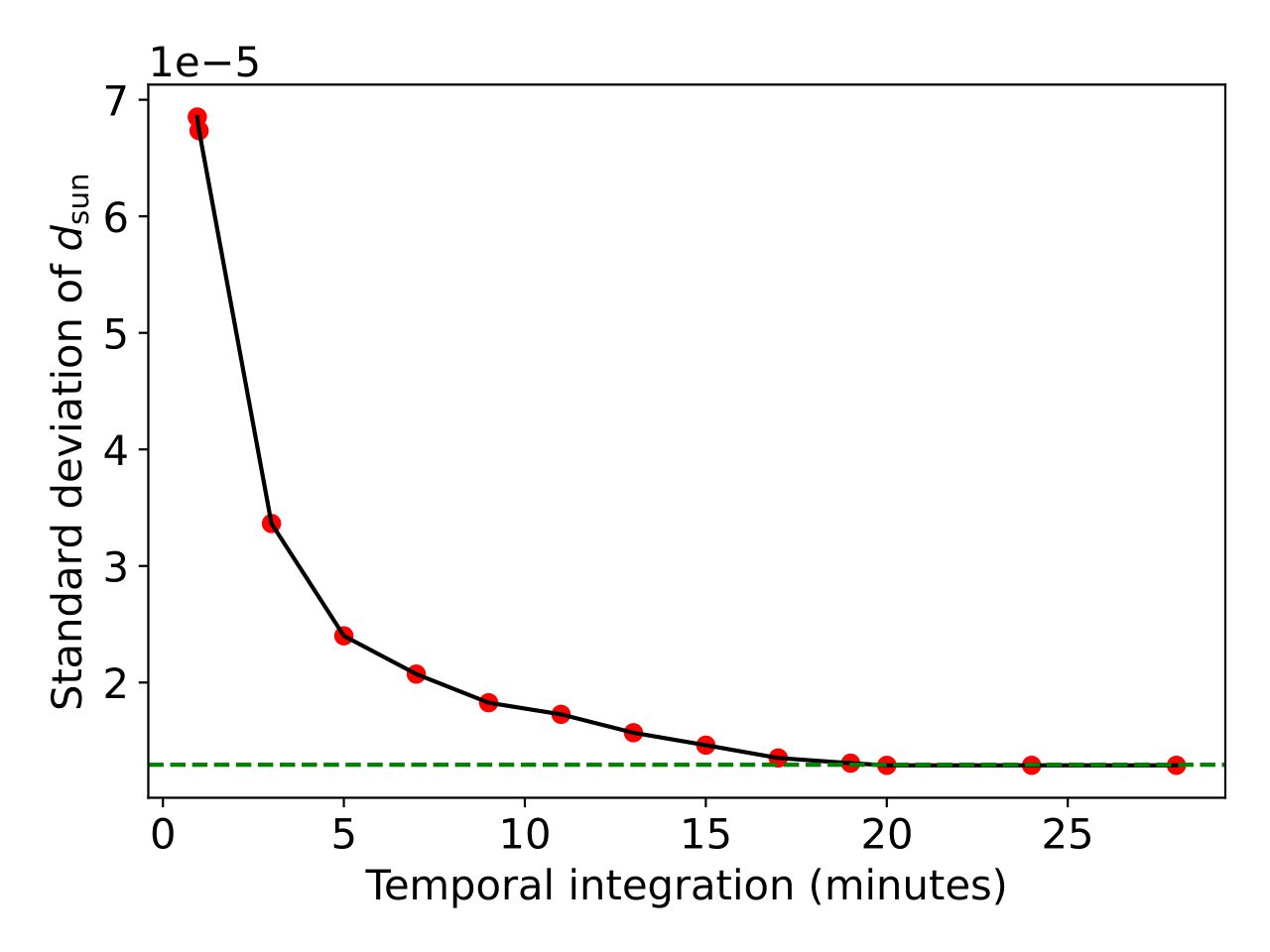}
    \caption{Left panel: Power levels increased due to noise diode firing during flux density calibrator scan, and increased by a reduced amount during the solar scan observed with the attenuation. Right panel: The variation of the band-averaged standard deviation of $d_\mathrm{sun}$ with temporal integration is shown. Beyond an averaging time of $\sim$15 minutes, the standard deviation of $d_\mathrm{sun}$ does not decrease further, indicating a saturation in the accuracy of the attenuation factor estimation (Reproduced from \citet{kansabanik2025solar}).}
    \label{fig:noise_power}
\end{figure}

Solar flux density calibration for MeerKAT relied on the use of noise diodes, which are located after the LNAs and before the attenuators in the signal chain \citep{kansabanik2025solar}.
The idea is straightforward and makes use of the fact that the noise diodes inject the same known amount of power in the signal chain, irrespective of the use of attenuators.
Let $P_\mathrm{cal,on}(\nu)$ and $P_\mathrm{cal,off}(\nu)$ represent the observed auto-correlation power with the noise diode on and off for a given spectral channel, respectively, on the calibrator source, as shown in the left part of the left panel of the Figure \ref{fig:noise_power}. Let us define this power difference on the calibrator, $d_\mathrm{cal}(\nu) = P_\mathrm{cal,on}(\nu) - P_\mathrm{cal,off}(\nu)$.
$P_\mathrm{sun,on}(\nu)$, $P_\mathrm{sun,off}(\nu)$ and $d_\mathrm{sun}(\nu)$ represent the same quantities for the Sun observed in the presence of attenuation, and hence, $d_\mathrm{sun}$ is reduced compared to $d_\mathrm{cal}$ as shown in the right part of the left panel of the Figure \ref{fig:noise_power}.

Then, the correction factor for flux density due to the use of attenuators is simply given by the ratio $d_\mathrm{cal}(\nu)/d_\mathrm{sun}(\nu)$ and also captures the spectral variation in the attenuator response.
For the case of MeerKAT the strength of the noise source in the UHF and L-bands is about half of the nominal system temperature, $T_\mathrm{sys}$. 
While this approach is sound in principle, it encounters a practical difficulty. 
Once the attenuation needed for solar observations has been applied (32-35$dB$, Sec \ref{subsec:analog-MeerKAT}), the step due to noise power injected by the noise diode becomes too small ($d_\mathrm{sun}$) to measure without integrating over a substantial time-bandwidth product.
The solar emission, on the other hand, is intrinsically not stationary in time or across frequency.
This leads to the consequence that the precision with which $d_\mathrm{cal}(\nu)/d_\mathrm{sun}(\nu)$ can be estimated does not continue to improve indefinitely with integration over a larger time-bandwidth product. 
When integrating over the entire MeerKAT UHF-band, it was found that this precision does not improve beyond an integration time of $\sim$15 minutes, as shown in the right panel of the Figure \ref{fig:noise_power}.

We note that for the SKA-Mid, the noise power of the diodes is expected to be smaller than that used for MeerKAT.
This implies that for estimating $d_\mathrm{cal}(\nu)/d_\mathrm{sun}(\nu)$ with SKA-Mid, a larger time-bandwidth integration will be needed to achieve a precision similar to that achieved with MeerKAT.
On the other hand, MeerKAT experience shows that the non-stationarity of solar emission limits the integration times to $\sim$15 minutes.

This implies that achieving high accuracy flux density calibration with the SKA-Mid will require precise characterisation of the attenuator response. 
These attenuators are similar to those are already in use at MeerKAT and have been characterised in some detail.
This included examining aspects like their spectral response, precision, the impact of attenuators on the observed visibility phases, and antenna-to-antenna variations due to manufacturing tolerances \citep{kansabanik2025solar}. 
It was found that while the spectral response of the attenuators was not flat, it deviated from the expected value by  $<$1~$dB$. The antenna-to-antenna variation in attenuator response was estimated to lie in the range of $\pm$2\% of the mean. 

\section{Calibration and imaging strategies for solar observations}\label{sec:cal_imaging} 
Given their large number of elements, the SKA telescopes and their precursors and pathfinders are able to achieve a good {\em uv}-sampling sufficient to characterize the solar morphology in unprecedented detail simultaneously along the axes of time, frequency, and polarization, and even when operating at a high temporal, spectral, and angular resolution. 
Naturally, this is accompanied by an enormous increase in data volumes and computational load.
Handling this efficiently will require a robust calibration and imaging pipeline capable of dealing with the entire range of solar conditions -- from very quiet to violently active -- without human supervision.
The snapshot spectropolarimetric solar imaging requirement implies that the optimization required for this use case is different from what is needed for the usual synthesis imaging case. 
Additionally, unlike the synthesis imaging case, snapshot spectropolarimetric imaging also does not lead to as dramatic a reduction in data volumes.
There are certain other differences that need to be accommodated as well. These include -- the calibration process needs to incorporate the instrument-specific procedure for flux density calibration; the extreme strength of the solar signal implies that even low-level calibration errors can become conspicuous; and the Sun is among the few astronomical sources whose emission is elliptically polarised -- can have large circular as well as linear polarization fraction \citep{Dey2025}.
The fact that the solar flux density usually dominates the entire FoV by a large amount also leads to a considerable simplification at SKA-Low frequencies, effectively reducing the imaging problem to the small FoV case. 

\subsection{SKA-Low precursors and pathfinders -- MWA and LOFAR} 

\subsubsection{Calibration and imaging of MWA solar observations}
The MWA solar observations are generally done by spreading 24 coarse channel bands, each of 1.28 MHz, over 72-300 MHz. The legacy MWA correlator allowed observations at 10 kHz/0.5s spectro-temporal resolution. The new MWAX correlator \citep{morrison2023mwax} allows observations at 0.25s time resolution, and the availability of fringe stopping in MWAX has enabled frequency averaging.

At the MWA, a standard calibrator is observed with the same spectral settings and attenuation as used for solar observations before sunrise and after sunset. These calibrator observations are used to derive the 4$\times$4 Jones matrices for full-polarimetric response of the instrument following the formalism described in \citet{Kansabanik2025-calibration-formalism}. This formalism uses an apparent-sky model with multiple linearly independent polarized sources to derive the instrumental Jones matrices. The polarized apparent-sky model is derived from the unpolarised sky model provided by GaLactic and Extragalactic All-sky MWA survey \citep[GLEAM;][]{Hurley2017} and the instrumental beam model of the MWA beam \citep{Sokowlski2017}. These instrumental Jones solutions are then transferred to the solar observations.

However, applying the calibration solutions from nighttime calibrator observations is not sufficient to obtain high dynamic range solar images, for which one needs to employ self-calibration. Since solar emission varies over small spectral and temporal chunks, spectroscopic snapshot full-polarisation self-calibration is performed following the algorithm described in \citet{Mondal2019} and \citet{Kansabanik2022_paircarsI}. This full-polar self-calibration not only improves the imaging dynamic range, but also reduces residual instrumental polarisation leakages based on an image plane correction assuming that the quiet Sun thermal emission is unpolarised. These self-calibrations are performed for individual 1.28 MHz spectral chunks and at $\sim30$s temporal intervals independently and in parallel. Solutions are interpolated while applying to the entire dataset. The final calibrated data is then imaged using an embarrassingly parallel framework. This entire calibration and imaging framework is implemented as an end-to-end pipeline - Polarimetry using Automated Imaging Routine for the Compact Arrays for the Radio Sun \citep[P-AIRCARS;][]{Kansabanik_paircars_2}.

\subsubsection{Calibration and imaging of the LOFAR solar observations} 
The LOFAR array includes 38 stations in the Netherlands, along with 14 international stations located across Europe. A group of 24 stations is located within an about 2~$km$ region in Exloo, the Netherlands, and is referred to as the core. At each station, there are two co-located antenna fields -- one of Low Band Antennas (LBA; 10--90 MHz) and two of High-Band Antennas (HBA; 110--240 MHz), providing 276 and 1128 instantaneous core baselines with the LBA and the HBA, respectively. LOFAR supports multiple simultaneous station beams, which enables concurrent observations of the Sun and a strong calibrator (for instance, a bright ``A-team” source), a capability commonly used for solar observing.

\begin{figure}[!htbp]
    \centering
    \includegraphics[trim={0cm 0cm 10cm 0cm},clip,width=0.8\linewidth]{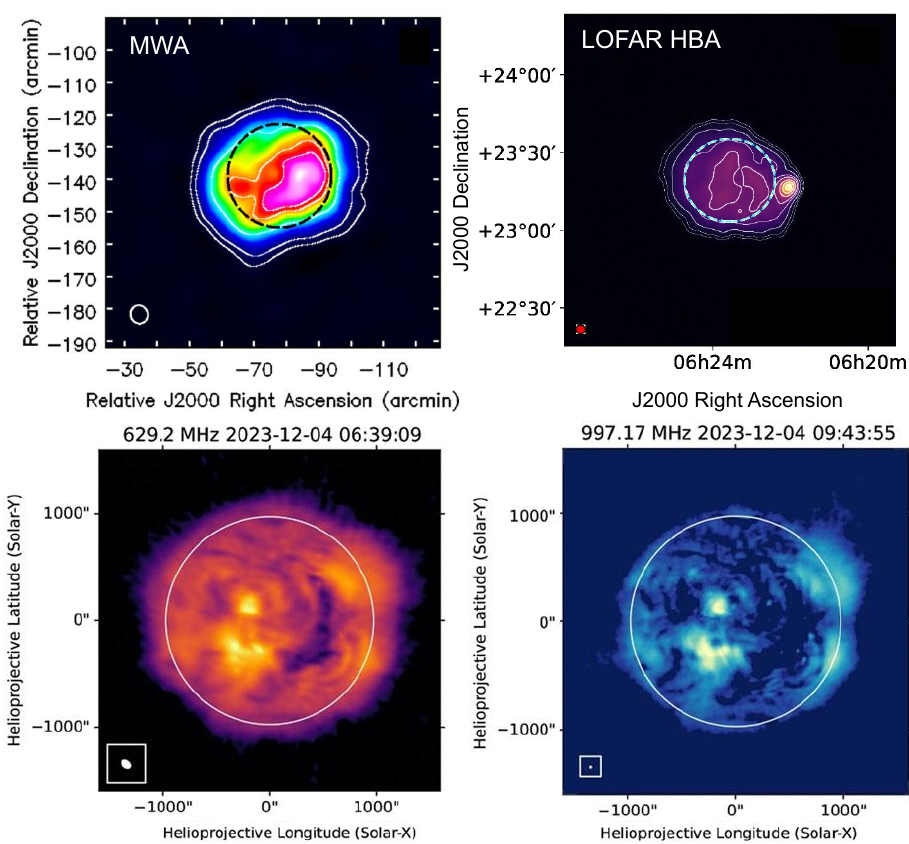}
    \caption{Top panels: High–dynamic-range spectroscopic snapshot images of the quiet Sun obtained with MWA and LOFAR are shown. The dotted disks mark the optical solar disk. The left panel shows an MWA image formed with 160 kHz bandwidth and 0.5 s integration using the Stokes I implementation of \textsf{P-AIRCARS} (\textsf{AIRCARS}), achieving a dynamic range of $\sim$1000 \citep{Mondal2019}. The right panel shows a LOFAR HBA image formed with 195.3 kHz bandwidth and 1 s integration using \textsf{SIMPL}, achieving a dynamic range of 544 \citep{dey2025automated}. Bottom panels: A demonstration of high-fidelity MeerKAT solar images made using 50 MHz and 15 minutes of spectro-temporal integration. The left image is from the UHF band, and the right image is from the L-band boresight observations of the Sun (Reproduced from \citet{kansabanik2025solar}).}
    \label{fig:paircars_simpl_meersolar_demo}
\end{figure}

Solar observations with LOFAR share all the calibration challenges that arise from the Sun’s extreme brightness and the wide field nature of aperture-array stations \citep{dey2025automated}. 
On the one hand, daytime calibrator beams are contaminated by solar emission, even when they are located several degrees away, and on the other, using calibrators far from the Sun can lead to large differential, direction-dependent ionospheric phase errors.
These two effects make a straightforward transfer of full complex gains from simultaneous calibrator observations unreliable for the direction of the Sun and motivate solar-specific calibration strategies.

To address these difficulties, the recent LOFAR solar calibration and imaging pipeline, Solar Imaging Pipeline for LOFAR \citep[SIMPL;][]{dey2025automated}, has adopted a hybrid approach that preserves calibrator-derived amplitude scaling but derives the phase solutions from the solar data itself via self-calibration. 
In practice, the pipeline first identifies the interval in the calibrator scan that is least contaminated by solar emission using statistical metrics on the calibrator dynamic spectrum described in detail in \citet{dey2025automated}, derives robust amplitude gains for absolute flux scaling, and then applies only the amplitude component to the solar dataset. 
The solar data are split into time–frequency chunks; within each chunk, an iterative self-calibration loop is run - starting from a simple initial model (typically a Gaussian at the phase centre using only those small baselines so that the Sun is unresolved) and progressively including longer baselines so that phase-only self-calibration converges before attempting amplitude-phase calibration. 
This strategy substantially improves imaging dynamic range and morphological fidelity compared to the transfer of full calibrator solutions. 

Two practical caveats remain. 
First, LOFAR’s remote baselines (out to $\sim100$ km) provide a very sparse sampling of the {\em uv}-plane, and the correlated solar flux typically falls off rapidly with baseline length, so the present high-fidelity solar imaging workflows generally concentrate on the dense core stations. Second, most operational pipelines for LOFAR solar imaging to date focus on Stokes-I imaging \citep{breitling2015, zhang2022, dey2025automated}. Both of these aspects -- incorporation of remote baselines and achieving full polarimetric calibration and imaging -- are priority developments to exploit LOFAR for high-resolution, spectro-polarimetric solar science fully.

\subsubsection{Demonstration of high-dynamic-range imaging}\label{subsec:demo_mwa_lofar}
Both \textsf{P-AIRACRS} and \textsf{SIMPL} are based on a similar philosophy of self-calibration rooted in their first incarnation, \textsf{AIRCARS} \citep{Mondal2019} and modified based on the instrument-specific needs and learnings gleaned over the years. Both of these pipelines lead to unprecedented high-dynamic-range spectroscopic snapshot solar imaging. Both can provide images with the dynamic range of several hundred for quiet Sun and exceeding several thousand during active emission. The top panels of the Figure \ref{fig:paircars_simpl_meersolar_demo} show two sample spectroscopic snapshot images of the quiet Sun from the MWA and LOFAR HBA observations made using the Stokes I counterpart of \textsf{P-AIRCARS} (\textsf{AIRCARS}) and \textsf{SIMPL}, respectively. This demonstrates that with a proper calibration strategy and imaging procedure these unsupervised pipelines can deliver science-ready high-fidelity spectroscopic snapshot solar images to the wider community, enabling diverse and rich science with the SKA-Low.

\subsection{SKA-mid precursor-- MeerKAT} \label{subsec:meerkat_calibraton}
Calibration of solar data with MeerKAT is more straightforward, when compared to that with the MWA and LOFAR. However, it does differ from the standard calibration for MeerKAT astronomical observations. At present, the solar observing mode has been validated for the UHF and L-bands of MeerKAT. MeerKAT solar observations are recorded with 4K channels and 2s time resolution in full-polar mode.  MeerKAT solar observations make use of standard flux density and phase calibrators, along with the noise diode for calibrating the solar attenuator used during solar observations (Sections \ref{subsec:analog-MeerKAT} and \ref{subsec:meerkat_fluxcal}). To make proper use of flux and phase calibrators, the following additional considerations need to be kept in mind. Pointing the telescope in the vicinity of the Sun leads to a significant increase in $T_{sys}$. 
A minimum zone of avoidance around the Sun has hence been defined for MeerKAT beams.
This minimum angular distance, $D_\mathrm{sun,min}$, is approximately $7^\circ$ for UHF and $4.5^\circ$ for L-bands (\href{https://archive-gw-1.kat.ac.za/public/meerkat/Solar-avoidance-radius.pdf}{MeerKAT Solar Avoidance Zone}). Therefore, calibrators or astronomical targets should only be observed at angular distances greater than $D_\mathrm{sun,min}$. 
Additionally, the passage of radiation through the turbulent solar wind introduces phase errors and scatter broadening, adversely impacting calibrator observations near the Sun.
The phase error due to such turbulence, $\phi_\mathrm{degree}$, can be estimated analytically (\href{https://library.nrao.edu/public/memos/vla/test/VLAT_236.pdf}{Butler 2005, NRAO}). For MeerKAT, assuming $\phi_\mathrm{degree} = 10^\circ$, yields the minimum angular distance from the Sun in degrees, $R_\mathrm{degree}$, of $\sim 15^\circ$ (UHF) and $\sim 10^\circ$ (L-band) \citep{kansabanik2025solar}.

As discussed in Sections \ref{subsec:analog-MeerKAT} and \ref{subsec:meerkat_fluxcal}, boresight solar observations at  MeerKAT frequencies are non-standard and demand specialized calibration and imaging strategies. To automate this calibration and imaging process, we developed a dedicated pipeline, \textsf{MeerSOLAR} \citep{kansabanik_2025_meersolar}, which is distributed through PyPI \href{https://pypi.org/project/meersolar/}{https://pypi.org/project/meersolar/}. This pipeline is fully automated, user-friendly, and deployable in both single-node workstations as well as high-performance cluster environments. It uses process-based parallelization using \textsf{Dask} \citep{dask-2015} and pipeline orchestration using \textsf{prefect}, enabling cross-platform execution. 

A master controller manages modular blocks, with independent tasks (e.g., attenuation calibration and data partitioning) running in parallel, while sequential tasks (e.g., standard calibration, self-calibration, imaging) are executed in order. Internal parallelism within blocks, such as per-scan calibration steps and time-chunked self-calibration, maximizes computational efficiency. In certain steps, there are some non-standard aspects to be considered, as listed below.  A detailed description of the pipeline is available in \citet{kansabanik2025solar}. 
\begin{itemize}
    \item {\textbf{Flagging:}} \textsf{MeerSOLAR} performs automated flagging on only flux and phase calibrators using \textsf{flagdata} task with \textsf{tfcrop} algorithm. Flagging is skipped for solar scans due to the intrinsic variability of solar emission.
    \item{\textbf{Basic calibration:}} To avoid solar contamination in calibrator observations, baselines $<200\lambda$ are flagged to avoid contamination from the large angular scale quiet-Sun emission. If contamination from a bright compact source on the Sun is still present, direction-dependent calibration is performed to remove the solar contribution.
    \item {\textbf{Imaging:}} In addition to its non-sidereal motion, the Sun exhibits differential rotation, 
    as discussed in more detail in Section \ref{subsec:solar-rotation}. 
    This leads to a solar latitude dependent rate of rotation. The largest projected motion of $\theta_{\mathrm{diff,rot}} \approx 5.0^{\prime\prime},\mathrm{hr}^{-1}$ occurs near the disk centre.
    To minimise image smearing due to this effect, integration times should be limited to $\sim$130 minutes in the UHF band and $\sim$50 minutes in the L band. 
\end{itemize}
A demonstration of the high-fidelity solar images obtained using MeerSOLAR for boresight MeerKAT solar observations at UHF and L-bands is shown in the bottom panels of the Figure~\ref{fig:paircars_simpl_meersolar_demo}. These images resemble closely the solar images in the extreme ultraviolet (EUV) band showing many common features between the two.
This demonstrates the ability of MeerSOLAR to produce high-fidelity solar images using MeerKAT. 
The algorithmic features of \textsf{MeerSOLAR} as well as its architecture, which prioritizes flexibility of implementation, are very pertinent for SKA-mid solar imaging. Its ability to handle data volumes not too different from what is expected from the SKA-Mid make it very interesting starting point for this application. 

\section{Other new-generation interferometers used for solar observations} 
\label{sec:other-instruments}

In addition to the instruments discussed in Sections \ref{sec:analog-signal-chain}, \ref{sec:flux-calib} and \ref{sec:cal_imaging}, there are multiple other new-generation instruments, including other SKA pathfinders, which are currently being used for solar imaging observations. 
The aspects relevant for observing the Sun with them are briefly described next. 
Table \ref{tab:instruments} summarises the key characteristics for all of these instruments.

\begin{sidewaystable}[htbp]
    \centering
    \caption{Your rotated table caption}
    \begin{tabular}{|c|c|c|c|c|c|c|c|}
    \hline
    \hline

        & Frequency & Baseline  & Angular  &  Time  & Frequency  & Nature of   & Observing   \\
        & range  &  range &  resolution\footnote{Corresponding to the longest baseline at the highest available frequency. The angular resolution practically achieved depends upon a variety of factors, ranging from elevation of the source to temporal and spectral span of observations and the weighting scheme employed.} &   resolution\footnote{Typical values} &  resolution\footnote{Typical values} &   instrument  &   since \\
        \hline
        & (MHz)  &  (m) &  (arcsec) &   (s) &  (kHz) &     &    \\

    \hline
    \hline
    LOFAR   & 10--90 & <100--10000\footnote{LOFAR offers baselines as long as 2000 km, however baselines beyond about 10 km are rarely used for solar observations}   & 84  &  0.16 &  195 &  Aperture array with   & 2010 \\
         &  &   &   &   &   & stations comprising clusters  &  \\
                  &  &   &   &   &   &  of dual-pol dipoles  &  \\
           & 110--240 &  <100--10000 & 31 & 0.16  & 195  & Aperture array with  & 2010 \\
            &  &   &   &   &   & stations comprising clusters &  \\
            &  &   &   &   &   &  of 4x4 dual-pol dipoles &  \\
    \hline
    NenuFAR & 10--85   & few--400  & 2220  &  1 & 3  & 96 elements each  &  \\
            &  &   &   &   &   & comprising an array of & 2019 \\
        &  &   &   &   &   &  19 dual-pol dipoles &  \\
    \hline
    OVRO-LWA &  12-85  & few--2500  & 355  &  0.2,10 & 96,24   &  Aperture array with & 2024 \\
        & & & & & & dual-pol dipoles &  \\    
    MWA     & 80--300 &  12--5300 & 47 & 0.25  & 10  & Aperture array with 256 & July 2013 \\
    &  &   &   &   &  & 4x4 dual-pol dipoles  &  \\

    \hline
    uGMRT   & 150--1450 & 100--25000  & 2  & 1.3 & 12.2 & 30 prime focus dishes & 2021 \\
    &  &   &   &   &  & of 45 m dia in Y config &  \\
    \hline
    DART    & 150--450  & 10--1000  & 168  &  1.7 & 2000  & 313 dishes of 6 m each  & Sept. 2023 \\
    &  &   &   &   &   & in a circular config  &  \\
    \hline
    MeerKAT\footnote{MeerKAT also offers S-band (170--3500 MHz), though it is yet to be commissioned for solar observations} & 580--1050 & 29--7700  & 9  &  2 & 16.602  & 64 offset Gregorian  & expected  \\
              & 900--1670 & 6  &   &   &  26.123 &  dishes of 13.5m dia  & in 2026  \\
    \hline
    EOVSA   &  1000--18000  & $\sim$10--1220  & 3 &  1 &  10000-40000 &  15 2.2m prime-focus dishes & 2017 \\
            &               &       & & & & and 1 27m cryogenically & \\
            & & & & & &  cooled dish for calibration & \\
    \hline

    \hline
    \hline
    \end{tabular}
    \label{tab:instruments}
\end{sidewaystable}

\subsection{Upgraded Giant Metrewave Radio Telescope (uGMRT)} 
\label{subsec:uGMRT} 
The upgraded Giant Metrewave Radio Telescope (uGMRT), a SKA pathfinder, operates in the 120--1450~$MHz$ range, which is divided into four observing bands:  120–250~$MHz$ (band-2), 250–500~$MHz$ (band-3), 550–850~$MHz$ (band-4), and 1050–1450~$MHz$ (band-5). The array comprises thirty antennas of 45~$m$ diameter each, arranged in a Y-shaped configuration, providing a maximum baseline of 25~$km$ and arcsecond angular resolution across its entire observing band. 

For reasons detailed in Section \ref{sec:analog-signal-chain}, the uGMRT needs to include substantial attenuation in the analog part of its signal chain to maintain it within its linear regime during solar observations.
The uGMRT design offers multiple options for introducing attenuation in the signal path -- in the radio frequency (RF) electronics at the front-end prior to converting the signal to optical signal for transport to the central electronics building (CEB); in the optical fibre (OF) system; and in the GMRT analog baseband (GAB), located after the signal is converted back to electrical in the CEB and before digitisation. The RF electronics provides coarse selectable attenuators of 14~$dB$ and 30~$dB$ in the front-end. One or both of these can be included in the signal path, yielding attenuations of 14, 30, or 44~$dB$. Both OF and GAB attenuators can provide attenuation of up to 30~$dB$ in steps of 0.5~$dB$ and 2.0~$dB$, respectively.
Some of the attenuation provided by OF and GAB systems is already utilized for keeping the entire signal chain at the optimal level for routine observations.


In practice, an appropriate combination of RF, OF, and GAB attenuation is used for different bands to ensure that the power levels entering the ADC remain within their linear regime during solar observations. As discussed in Section \ref{sec:flux-calib}, there is an additional challenge to deal with, that of absolute flux density calibration. This gets progressively worse at higher frequencies, where the disparity between the flux densities of calibrator sources and the Sun grows extremely large.
An additional tool that can be employed for flux density calibration is the use of the noise diodes, which inject a known amount of noise power in the signal path, as done for MeerKAT solar observations \citep{kansabanik2025solar}. The uGMRT noise diodes are designed to provide four selectable noise power levels.  
Standardized procedures for solar observations and flux density calibration have been established for bands 2 and 3, while analogous procedures for higher bands continue to be refined.

Despite these challenges, recent works using band-2 \citep{Mondal_2024_gmrt} and band-3 \citep{Mondal2025} have demonstrated the significant potential of uGMRT for niche solar studies, providing high angular resolution observations of active emissions at low frequencies. These studies detected compact radio sources from active solar emission, which is 2-3 times smaller than the prediction made based on the current scattering model of the solar corona.
\citet{Dey2025} have recently demonstrated full-Stokes solar imaging in the band 2. This observation made simultaneously with the MWA in overlapping spectral range makes the first robust detection of linearly polarised emission from solar active emission, an important step toward the meterwavelength solar polarimetry.

\subsection{DAocheng Radio Telescope (DART)}
\label{subsec:DART} 

DAocheng Radio Telescope (DART), located at at altitude of approximately 3820~$m$ in Daocheng, Sichuan, is a key component of China’s Meridian Project II. As the world’s first solar-dedicated aperture array combining synthesis imaging, ultra-broadband reception (150--450~$MHz$), and a dense ring configuration (313 $\times$ 6~$m$ parabolic antennas arranged on a 1~$km$ diameter ring), DART delivers transformative capabilities in low-frequency radio astronomy.

DART provides high fidelity snapshot spectropolarimetric solar imaging with angular resolution of $\sim$100" ($\sim$300") at  450~$MHz$ (150~$MHz$).
Its $>20^\circ$ FoV at 150\,MHz, enables extended coronal monitoring. 
It offers an instantaneous bandwidth of up to 40~$MHz$, covering the full 150-450~$MHz$ band in 8 steps, and also a ``picket-fence'' mode which samples multiple narrowband channels distributed across the full bandwidth for monitoring purposes.
DART offers flexible observing modes to accommodate a diverse set of science goals, providing time resolution ranging between 5--100~$ms$ and frequency resolution between 10~$kHz$ and 2~$MHz$.

DART’s pipeline enables 
imaging without iterative deconvolution for typical solar data. Using multi-scale CLEAN, it achieves a dynamic range of more than 4000. Initial polarization calibration shows cross-polarization leakage $<10\%$, with solar type I and III bursts exhibiting circular polarization up to 100\%, demonstrating robust polarimetric capability despite limited polarized calibrators in this band.

DART has already started delivering interesting science results across different areas of solar physics, ranging from the study of large numbers of type-I radio bursts; prediction of interplanetary shocks from tracking of type-II radio sources; tracing the cold and dense plasma in the flux rope of CMEs revealing 3D eruption dynamics before the CME is visible in white-light coronagraphs; and detection of pulsation in active region loops with 2-18 minutes periodicities observed simultaneously in bands across the electromagnetic spectrum allows us to link magnetohydrodynamic (MHD) waves to periodic magnetic reconnection. 
These results establish DART as a powerful pathfinder for SKA. 
DART provides critical technical and scientific lessons for future large arrays such as the SKA, particularly for solar observations. 
Early results from DART are now beginning to appear \citep[e.g.][]{Feng2025-DART} and DART data will be publicly released through a dedicated archive (planned for 2026). 

\subsection{New Extension in Nan\c{c}ay Upgrading LOFAR (NenuFAR)}
\label{subsec:NenuFAR}
The New Extension in Nan\c{c}ay Upgrading LOFAR \citep{Zarka2015,Zarka2018} is a new-generation low-frequency radio array comprising 96 mini-arrays (MAs, each consisting of 19 antennas) from the core within $\sim400$m and six distant MAs extending out to $\sim$2.5 km for improved imaging angular resolution. It operates in the 10-85 MHz band and provides beamforming as well as interferometric imaging observations of the Sun \citep{Briand2022}. Very high resolution (0.1 kHz and 0.3ms), high sensitivity dynamic spectra from beamforming observations at these very low frequencies are an asset, and provide a detailed sub-structural view of the well known radio bursts. The smaller and dense footprint of the NenuFAR provides high surface brightness sensitivity, though only a moderate resolution of $\sim6$arcmin, when using both core and distant MAs. This angular resolution is unmatched at very high spectro-temporal resolution it provides, making it very suitable for detailed spectroscopic snapshot imaging studies of compact solar radio bursts.

\subsection{Expanded Owens Valley Solar Array (EOVSA)}
\label{subsec:E-OVSA}
The Expanded Owens Valley Solar Array (EOVSA) is a solar-dedicated radio interferometer capable of observing the full Sun between 1--18 GHz at a cadence of 1s. A variety of calibration procedures are performed to ensure that the EOVSA data can be used to produce quality science. The key ones are listed below:
\begin{enumerate}
    \item \textbf{Flux calibration:} The fluxscale of EOVSA is currently based on the disk-integrated flux density measurements reported daily by the Radio Solar Telescope Network (RSTN). The fluxscale is determined every day by assigning the observed flux density when the Sun is at a high elevation and is not undergoing any flare, to that reported by the RSTN for that day.
    \item \textbf{Attenuator calibration}: To compensate for the very high solar flux density, EOVSA uses multiple attenuators, which ensure that the system is maintained in the linear regime and produces quality data. The attenuation levels are changed automatically in response to changes in the solar flux density, and the corresponding level and timestamp are recorded in a central database. This information is then used for flux density calibration. The attenuators themselves are calibrated periodically by measuring the flux density of the background sky, with the attenuators on and off.
    \item \textbf{Gain calibration}: To calibrate the temporal variability of the instrument, well modelled 
    calibrator sources, are observed. However, its 2.1~$m$ dishes, required to observe the full solar disc over the large frequency range, do not have the sensitivity to observe the weak calibrator sources. To overcome this, EOVSA uses a 27~$m$ dish with  cryogenically cooled receivers. Baselines between the 2.1~$m$ dishes and the 27~$m$ dish have sufficient sensitivity to observe calibrator sources over this large frequency range and measure the gain variations of the instrument. 
\end{enumerate}
EOVSA has already provided crucial insights into different areas of solar physics \citep[e.g.][]{fleishman2020,wei2021, chenbin2024}. While EOVSA records dual-polarisation data, and the intensity data products are currently available publicly\footnote{
https://www.ovsa.njit.edu/wiki/index.php/EOVSA\_Data\_Products}. 

EOVSA is currently being upgraded, which includes addition of two new antennas, as well as adding new feeds to all antennas which are optimised for the EOVSA dishes. These improvements are expected to make EOVSA capable of delivering images with much higher fidelity than possible earlier. The new feeds are also expected to provide accurate polarisation data as well. EOVSA results serve as demonstrations of the science results enabled by high-frequency spectroscopic snapshot solar imaging, which the higher frequency bands of the SKA-Mid will enable. 

\subsection{Owens Valley Radio Observatory Long Wavelength Array}
\label{subsec:LWA}
The Owens Valley Radio Observatory Long Wavelength Array (OVRO-LWA) is a newly commissioned low-frequency radio interferometer comprising 352 antenna elements with a maximum baseline of 1.5 km and operating in the 15-88 MHz frequency range. It is an all-sky imager and hence observes the Sun 
practically, whenever the Sun is above the horizon. 
Stokes I calibration and imaging pipeline of OVRO-LWA has been released (\href{https://pypi.org/project/ovrolwasolar/}{available through PyPI}), and is presently being used to produce solar images in a semi-realtime mode\footnote{
https://www.ovsa.njit.edu/lwa}. The currently operational pipeline continuously produces images at a frequency resolution of 312~$kHz$ between 32--88~$MHz$ at a cadence of 20~$s$. It also produces images with a frequency resolution of approximately 5 MHz within the same frequency band at the same time cadence. Polarisation calibration and imaging pipeline are under development. By default, the data is recorded with zenith as the phase center. During processing, the pipeline performs a self-calibration step using a sky model, which is generated from the data itself. Next, it subtracts all sources outside a region centred on the Sun with radius approximately $10^\circ$. Then the phase center is rotated to the solar centre, and imaging is done. It is found that despite centring on the solar centre, the apparent location of the Sun gets shifted due to ionospheric refraction. Hence, an additional ionospheric shift correction needs to be applied. While a shift-correction algorithm has been implemented in the currently operational pipeline, it might not be suitable for all science cases.

\section{Other solar-specific considerations}

The significant differences between the characteristics of the Sun and its radio emissions and what most sensitive interferometers are optimized for imply that, in addition to what has already been discussed, there are multiple other considerations that come into play when optimizing these instruments for solar observations. The most important of them are discussed briefly next, and also include recommendations for the SKA telescopes where relevant.

\subsection{Self-noise considerations}
\label{subsec:selfnoise}
Solar flux density is large enough to dominate the system noise of individual antennas across most, if not all, of the SKA frequency range. Even if it does not dominate the system noise, due to the large number of antennas, the total solar flux intercepted by the array ($NS$, where $N$ and $S$ are the total number of antennas and solar flux density, respectively) can become larger than the system noise of a single antenna. Both these conditions make the system fall in the so-called `strong noise regime'. The theoretical background of the strong noise regime and its effects on calibration and imaging are detailed in seminal works like \citet{kulkarni1989} and \citet{vivekanand1991}. The effect of strong noise on solar images is discussed in detail by \citet{bastian2025a, bastian2025b}. Here we shall only note some key issues relevant for solar imaging and calibration.

\begin{enumerate}
    \item The noise on each visibility measurement depends not only the system temperature of the two antennas comprising the baseline, but also on the flux intercepted by the baseline. Solar flux density can, however, have a very strong dependence on baseline length. This implies that some changes might be needed in the calibration methods implemented in standard softwares, as they generally assume that the uncertainty on each visibility measurement depend only on the system temperature, and hence have similar noise associated with them.
    \item Off-source noise in solar images is limited by the system noise only if deconvolution is perfect and the integration is done over a very small time and frequency range \citep{bastian2025b}\footnote{In reality, this only happens if total power measurements are available. As shown in \citet{bastian2025a}, for large N-arrays like SKA, the noise asymptotically approaches the noise achieved when total power measurements are used.} If these conditions are not met, then the off-source noise is also dependent on the source noise.
    \item In the case of a point source, the on-source noise is independent of the number of available baselines, and is only dependent on the frequency, time integration, as well as the source flux. In absence of such a strong point source on the Sun, the on-source noise also depends on the uv-coverage and the source brightness distribution as well.
    \item For beamforming measurements, since the total power is also added, the noise in the measurements will in general independent of the instrumental noise, but only depends on the frequency and time integration.
\end{enumerate}

\subsection{Elastic processing} 
\label{subsec:tradeoffs}

The phrase {\em elastic processing} has been borrowed from the field of computing models, where it refers to automatic scaling of resources to match demand. 
In the present context, one would ideally want to match the time and frequency resolution of the solar radio images, and hence the visibilities, to the temporal and spectral variability scales of the solar emission under study.

For the Sun, these variability scales span many orders of magnitude: from millisecond, kilohertz–wide structures to quasi-static quiet-Sun emission over minutes and $\gtrsim 10$~MHz-wide bands. Modern large-$N$ correlators, however, are typically constrained by data-distribution architectures and internal throughput. In practice, the shortest integration times they can provide at full bandwidth and full $uv$-coverage often fall well short of what would be ideal for the most rapidly varying solar phenomena.

\begin{table*}[!htbp]
    \centering
    \small
    \renewcommand{\arraystretch}{1.2}
    \begin{tabular}{p{0.12\textwidth} p{0.22\textwidth} p{0.23\textwidth} p{0.30\textwidth}}
        \hline
        Mode & Example science case & Typical $\Delta t$, $\Delta f$ & Primary trade-off \\
        \hline
        Burst mode 
        & S-bursts, type~IIIb/I  
        & $\sim$ms, $\sim$kHz 
        & Reduced number of baselines and sparsely sampled subbands \\
        
        Active mode  
        & Type~II / type~III bursts 
        & $\sim$0.1--1~s, $\sim$100~kHz 
        & Moderate reduction in correlated bandwidth and/or baselines. \\
        
        Regular mode 
        & Quiet Sun, CMEs, long-term monitoring 
        & $\sim$10~s, $\sim$100~kHz 
        & Coarser $\Delta t$, $\Delta f$ and/or sparse band to minimise data rates and storage. \\
        \hline
    \end{tabular}
    \caption{Illustrative elastic-processing modes for solar observations, highlighting the trade-off between time resolution, spectral resolution, correlated bandwidth, and effective correlator throughput. By adapting the mode to the variability level, one can obtain detailed coverage of rapidly evolving events while remaining data-rate and storage-friendly during quiet periods.}
    \label{tab:elastic_modes}
\end{table*}

A practical way to overcome this limitation is to trade spectral resolution, correlated bandwidth, and/or the number of baselines against time resolution while respecting the data-distribution and throughput constraints. 
For solar imaging, additional flexibility is available as accepting a substantially sparser UV-coverage and lower correlated bandwidth can still yield a scientifically more attractive data product if it offers images at much higher time resolution.  
For SKA-Low, an especially attractive option is to reduce the {\it effective} bandwidth by using a sparse ``picket-fence'' spectral sampling: for example, $\sim 100$ narrowly correlated channels (each $\sim100$~kHz wide) distributed across the full 50–350~MHz band, for a total correlated bandwidth of only $\sim 10$~MHz, but still retaining the global spectral context.

Elastic processing formalises these trade-offs into a small number of observing modes, each tuned to a characteristic variability regime. In a ``burst'' mode, the correlator can use only a subset of baselines and a sparsely sampled subband in order to deliver millisecond cadence and kilohertz channels for fine-structure studies. In an ``active'' mode, appropriate for typical type~II/III activity, integration times of order 100s of $ms$ and channel widths of order 100~kHz suffice, with moderate reductions in correlated bandwidth or baseline set. During quiet periods, a ``regular'' mode with coarse time and frequency sampling, possibly combined with a picket-fence band layout, provides adequate science return while dramatically reducing data rates and storage requirements. Table~\ref{tab:elastic_modes} summarises these illustrative modes. While a $\sim 100~kHz$ channel resolution is expected to work well for most science cases, we note here that the channel resolution will determine the effective field-of-view of the instrument, which will not be significantly affected by bandwidth smearing and will depend on the longest baseline available for an observation. For example, for achieving an angular resolution of $5''$ over a $1^\circ$ field-of-view at 200 MHz, 100 kHz is the maximum channel resolution which can be used.

Even when not fully automated, such elastic processing offers a powerful way to concentrate correlator and storage resources where they matter most: the instrument can operate in a storage-friendly regular (and possibly sparsely sampled) mode by default, and switch to higher-cadence modes when solar activity is detected (Section \ref{subsec:triggered-obs}) or anticipated, enabling much more detailed processing of transient events without overwhelming the system.


\subsection{Need for a dedicated solar imager}
\label{subsec:SDP-solar-mode}
The default data product from the SKA science data processor (SDP) is expected to be an image cube of the requisite time and frequency resolution. While there is likely to be some flexibility in choosing different time and frequency resolutions for the data cube, the bulk of the SDP imaging pipeline is optimized for generating image cubes over long time integrations. The visibility data over the entire duration is expected to be used for deconvolution, which is expected to produce very high fidelity and high dynamic range radio images. Another imaging mode suitable for transient science is also expected to be available in the SDP. This involves two steps, with the first step being the subtraction of the continuum model sky visibilities, and then using the residual visibilities to produce image cubes at high time and frequency resolution. However, these image cubes will not undergo any deconvolution. Unfortunately, none of these modes is suitable for solar imaging, as it's requirements are not met by these SDP modes. We note that:
\begin{enumerate}
    \item The radio Sun is extremely dynamic and shows a lot of spectro-temporal features. Hence, detailed investigations necessarily require deconvolving visibilities at high time and frequency resolution. The transient imaging mode of SDP is not suitable for this purpose -- there is no good sky model to subtract and there are often multiple dynamic sources simultaneously on the solar disc. 
    \item The accuracy requirement of the primary beam model  for high fidelity full Stokes solar imaging is very high. This is because of solar flux density as well as the polarisation fraction of solar emission can be very high,  and small errors in the beam model, which might otherwise be buried in noise, become noticeable when observing the Sun.
    \item There is also the additional complication due to solar differential rotation. Discussed in detail in Section \ref{subsec:solar-rotation}, this limits high angular resolution solar imaging to short integration times and also breaks the ``rigid-sky" assumption routinely made for interferometric imaging. 
\end{enumerate}

\subsection{The need for triggered observations} 
\label{subsec:triggered-obs} 
The SKA telescopes are expected to be significantly oversubscribed, and the solar science case, like all other science cases, will need to compete for precious observing time. 
While some of the most interesting times for solar observations are during periods of solar activity \citep[as discussed in multiple articles in this volume, e.g.][]{Kontar.1.2026.SKA, Kumari.1.2026.SKA, Morosan.1.2026.SKA}, the episodes of solar transient activity are inherently unpredictable. This makes blind scheduled observation very inefficient at capturing solar activity and very prone to missing big solar events. It also leads to a major bottleneck in building a statistically large ensemble of imaging studies of different kinds of solar activity. In order to make optimal use of the limited solar observing time, one must implement a triggered solar observing mode. 
Triggered solar observations are also discussed in some detail by \citet{GemmaAnderson01.2026.SKA} in the present volume.

Recently, such a triggering framework relying on an external trigger has been implemented at the MWA \citep{patra_stormy}. It has already captured about 110 bursts over a period of 6 months, demonstrating the efficacy as well as efficiency of such trigger-based observations. In this specific case, this framework uses unallocated observing time during the day, in conjunction with the MWA voltage buffer capture system \citep{morrison2023mwax}, which can store up to the past $\sim$180s worth of data. This ability allows the MWA to partially mitigate the data loss due to the latency of $\sim$4-minutes between the event-associated radio emission arriving at Earth and the arrival of the trigger at the MWA. The bulk of this latency arises due to the use of triggering from an external solar-dedicated spectrograph, located at Yamagawa in Japan \citep{Iwai_2017}, and can be reduced substantially by using an internal trigger.

Here, we confine ourselves only to the aspects related to the generation of triggers for observing solar activity. The aspects related to initiating observations with the SKA telescopes, in response to these triggers, are beyond the scope of the present discussion.
A few different observing strategies can fruitfully be explored for triggered solar observation using SKA telescopes.  First, a range of approaches can be taken for real-time trigger generation. For example, a wide-band co-located solar spectrograph can provide an external trigger for SKA telescopes. Alternatively, or perhaps in parallel, data from suitably chosen SKA-Low stations and SKA-Mid dishes can itself be used for generating internal triggers. Both SKA telescopes have sub-array capabilities. SKA-Low also provides substation capability as well as multiple substation beams. Using small subarrays to monitor the Sun has also been suggested \citep{seethapuram_sridhar_2025_16951088} for both SKA-Low and SKA-Mid. For the SKA-mid, they suggest using a set of 3 subarrays, of 4 antennas each, to simultaneously monitor solar activity in the 3 SKA-mid bands. Such subarrays can provide data that can be used to track and detect solar activity in real-time and generate observing triggers for the entire array as appropriate. 

We also suggest an alternative approach, suitable for daytime commensal observations, with the SKA-Low, making use of its substation capability.
The smallest substation currently defined has a footprint of 6~$m$ \citep{SKAO-Low-SubStation-Templates}. A subarray comprising these small substations of $\sim$60 SKA-Low stations within a 10~$km$ diameter, adds up to $<$0.5\% of the collecting area, and thus has minimal sensitivity impact on any commensal observations. A sparse spectral sampling (e.g. 100 channels of width 5.4~$kHz$ spread across 50--350~$MHz$) is quite sufficient for solar monitoring, and uses a tiny fraction of the telescope resources. In addition to a flexible spectral configuration, SKA-Low also provides a substantial transient buffer, spanning 900~$s$ \citep{SKAO-YearInLife}. If data from the solar monitoring subarray can be directed to the buffer, it will enable the possibility of retrieving event data before the initiation of the triggered observations and mitigate the loss of event data due to the time taken for initiating the triggered observations. Additionally, these data only need to be recorded when an event is detected, and a trigger is raised. 

The radio-based triggering system will only capture events with radio signatures bright enough to be detected in near real-time, typically using non-imaging methods. 
However, there are several flare and CME events, which are of high space weather relevance, but do not produce dominating radio emission. Hence, in conjunction with radio-based triggering, triggering based on near-real-time space-based instruments is also needed. At present, space-based instruments provide CME triggers with 1-5 hours of latency, not sufficient for near-Sun triggered observations. An upcoming SunCET \citep{Mason2021} EUV cubesat mission will provide much faster CME detection alert using UHF beacons using on-board non-imaging techniques. An integrated triggering based on the SunCET or similar missions during the SKA era will play a crucial role. 

\subsection{Usefulness of access to a small-subset of visibilities}
\label{subsec:preserve-visibilities}
While it is clear that it will not be feasible for the SKA to save visibilities, nonetheless, for solar observations, where even individual visibilities can potentially provide high SNR measurements, there are some advantages to recording a small fraction of visibilities, which then do not need to be imaged in near real time.
One can potentially explore approaches like recording $\sim$1\% of the visbilities to limit their storage and analysis impact. The axis along which to thin the sampling depends upon the nature of solar emission under consideration. Most often it will be spectral axis which along which one can tolerate sparse sampling, so this might simply translate to saving $\sim$1\% of the spectral channels.

\subsection{Distinguishing between RFI and active solar emissions}
\label{subsec:RFI-flagging} 

RFI refers to all signals of non-astronomical origin that contaminate the astronomical signal. Mitigating RFI is a crucial part of radio astronomy data processing for practically all radio facilities. Though great care has been taken to choose the site of SKA instruments in parts of the globe with extremely low population densities and inside extended and well-enforced {\it Radio Quiet Zones}, RFI mitigation will still be required for SKA data. Over the years, a diverse set of algorithms to deal with RFI, both during post-processing and in real time, has been developed. 

Flagging of RFI contaminated data during post-processing is the most common RFI mitigation step after correlation. Existing automated algorithms in the community standard radio data processing software like CASA \citep{mcmullin2007casa} and AOFlagger \citep{aoflagger} identify and flag RFI in the time–frequency plane, typically using a threshold and outlier detection approaches.
They implement sophisticated approaches for aiding outlier detection, including methods like frequency/time smoothing and morphological dilation/scale-invariant-rank operators to pick up different kinds of RFI. 
More recently, machine-learning techniques (CNN, deep U-Nets, etc.) have been applied to classify RFI in spectrograms \citep[for e.g.][]{akeret2017radio, yang_FAST,sun_cnn,du2024comparison}.
Apart from post-processing, RFI mitigation can also be done in real time. Dedicated hardware (FPGAs/GPUs) can detect and excise RFI on-the-fly. For example, spectral-kurtosis (SK) estimators, which flag non-Gaussian fluctuations \citep{nita_rfi}. Another example from  \citet{Buch2019JAI} implements an approach tailored to deal with impulsive RFI and is in use at the uGMRT. Recent work by \citet{buch2025real} provides an extensive survey of commissioned and emerging RFI-mitigation methods, including excision algorithms, analog suppression schemes, and AI/ML-based approaches. In practice, a combination of both real-time and post-processing approaches is used for RFI mitigation.

However, these methods focus mainly on separating weak, stationary astronomical signals from RFI affected data. In case of solar observations, the active solar emissions are highly time and frequency variable, can be much stronger than the background solar emission, and also follow non-Gaussian statistics.
Genuine solar radio bursts can therefore easily be mistaken for RFI \citep{Kansabanik_paircars_2,dey2025automated}.
Solar bursts satisfy most, if not all, of the characteristics typically used to identify RFI by mitigation algorithms, rendering them unsuitable for solar observations. 
These limitations have been successfully overcome for the MWA solar observations by using approaches that look for outliers in the uv-plane for every single spectro-temporal chunk based on the smoothness property of the visibility distribution of any true astronomical source. This new flagger, named aNKflag \citep{Kansabanik_paircars_2}, has been used in P-AIRCARS. However, this method does not apply to single-dish or beamforming observations. 

There have been multiple rigorous efforts in identifying solar radio bursts in archival dynamic spectrum using a variety of different approaches, including deep learning/ML based methods, which rely on using prior knowledge about the nature of specific kinds of solar radio bursts to distinguish them from RFI \citep[e.g][]{Lobzinn,salmane,scully2021type,orue2023automatic,deng2024real}. However, efforts that deal in real-time simultaneously with RFI mitigation and solar burst detection have only just begun to appear \citep{patra_stormy}. These real-time flaggers are essential for a dynamic spectrum-based triggering system.  


The various aspects of the problem of flagging RFI in the presence of solar activity are now fairly well understood, and considerable progress has been made in reliable automated separation of RFI from a large variety of active solar emissions. 
With improved adoption and optimization of ML-based tools as this work progresses with SKA precursors and pathfinders, as well as other instruments, it is reasonable to expect to have robust real-time as well as post-processing tools for mitigation of RFI during solar observations. 
We note that the comparatively low RFI environments of SKA telescopes are expected to reduce the complexity of the problem significantly.


\subsection{Dealing with differential solar rotation} 
\label{subsec:solar-rotation}
Like all solar system objects, the Sun is a non-sidereal source. The non-sidereal motion of the Sun is routinely accounted for while observing and correlating, but is usually limited to tracking the centre of the Sun. 
The Sun also rotates on its own axis, and unlike a rigid body, this rotation is differential in nature, with rotation periods of $\sim$25 days at the equator and $\sim$34 days near the poles \citep{Mancuso2020}. 
As a result, solar features at different latitudes move across the plane of the sky at different rates, with the largest projected motion occurring near the disk centre. The maximum differential rotational motion is $\theta_{\mathrm{diff,rot}} \approx 5.0^{\prime\prime}/\mathrm{hr}$. 
Not accounting for these issues adequately could lead to complicated smearing effects in the image plane. 

One approach to deal with this issue is to make images at a sufficiently high cadence, such that the solar rotation is much smaller than the instrumental spatial resolution, followed by correcting for the solar differential rotation in the image plane, before stacking these images to generate the final synthesis image. However, this method needs imaging at very high cadence at higher frequency bands of SKA-Mid 
but might lead to low fidelity of deconvolution, especially for the fainter sources. This would, in turn, lead to errors in the final synthesis image. A better approach would be to explore ways to account for the solar differential rotation during the deconvolution step. We note that differential solar rotation breaks the ``rigid-sky" assumption of radio interferometric imaging, making corrections in the visibility domain or during the imaging and deconvolution non-trivial. Current tools like Common Astronomy Software Application \citep[CASA;][]{CASA2022} and W-Stacking CLEAN \citep[WSClean;][]{Offringa2014} do not support such corrections. The community needs to develop tools for this purpose to fully make use of the extremely sensitive high-angular-resolution observations available with the SKA, particularly the SKA-Mid.

\section{Considerations specific to heliospheric observations}
\label{sec:helio-considerations}
Ground-based radio observations can directly detect emission from solar plasma out to heliocentric distances of $\sim15\ R_\odot$. Beyond this range, the heliospheric plasma is probed indirectly using the propagation effects it imprints on the emission from the background radio sources.
This is done using techniques like interplanetary scintillation (IPS), Faraday rotation (FR), and angular broadening measurements. The scientific importance of these approaches is discussed in detail in other chapters of this volume \citep{Chhetri.1.2026.SKA, Chrysaphi.1.2026.SKA, Dey.1.2026.SKA, Kansabanik.1.2026.SKA, Nakariakov.1.2026.SKA, Raja.1.2026.SKA, P.Zhang.1.2026.SKA}. Although these methods rely on observations of astronomical sources, achieving the heliospheric science goals with both SKA telescopes requires careful consideration of observational strategies, calibration procedures, and imaging methodologies. This section outlines these requirements and summarise the current status of realising these objectives.

\subsection{Observing strategy}\label{subsec:helio-obs-strategy}
Detecting faint background radio sources ($\lesssim$few Jy) near the Sun, whose emission can exceed $10^4$ Jy, demands careful mitigation of solar contamination. The optimal strategy is to place the Sun at a primary-beam null, but strong beam chromaticity makes this difficult across wide bandwidths, even though heliospheric studies require instantaneous broadband coverage. This conflict can be resolved by exploiting the large number of SKA antennas and dividing the array into multiple sub-arrays. Each sub-array observes a narrower frequency range (within SKA-Low or a single SKA-Mid band), over which the Sun can be consistently placed at the beam null while preserving adequate sensitivity. Operating both SKA telescopes in this mode enables simultaneous heliospheric measurements from the outer corona to the inner heliosphere with near-nominal system sensitivity.

Heliospheric observing requires two distinct scheduling modes tailored to different science goals. For studies of background solar-wind plasma and relatively stable heliospheric structures such as stream or corotating interaction regions (SIRs/CIRs), observations can be planned in advance, analogous to standard astronomical scheduling, while accounting for the relevant pointing constraints.

A second class of observations targets transient phenomena such as CMEs. CME onsets, propagation directions, speeds, and angular extents cannot be predicted in advance, making blind heliospheric observing inefficient. To address this, we have developed a triggering system, ``helioschedule", which enables CME-driven heliospheric observations. The system uses near–real-time coronagraph data to estimate optimal telescope pointings and observing windows, and triggers observations accordingly. A detailed description is provided in another chapter of this volume (Kansabanik et al.). Implemented and operational on SKA precursor and pathfinder facilities such as the MWA and ASKAP since August 2024, helioschedule has demonstrated high efficiency in capturing CMEs while optimising telescope time. A comparable triggering capability will be essential for heliospheric observations with the SKA.

\subsection{Adaptive field of view for SKA-low} \label{subsec:adaptive-fov}
As a CME erupts and propagates through the heliosphere, it undergoes continuous expansion. Close to the Sun, instruments with a small FoV are well suited, as a single pointing can cover a large fraction of the CME while keeping the Sun at a primary-beam null. At larger heliocentric distances, however, multiple pointings are required to image the expanding CME with such narrow FoVs. Conversely, wide-FoV instruments are better suited at larger distances, where they can capture a substantial portion of the CME in a single observation. These instruments are ineffective close to the Sun, since the Sun enters the FoV, increasing the system temperature and thereby reducing sensitivity.

Hence, an instrument with an adaptive field of view (FoV) is well suited to meet both requirements. The full SKA-Low station beam provides an FoV of $\sim3-4^\circ$ ($\sim10\,R_\odot$) at the centre frequency, which is optimal for observations in the outer corona. However, probing the inner heliosphere requires FoVs of several tens of degrees. For instance, the MWA, with its very wide FoV ($\sim30 \times 30~\mathrm{deg}^2$), can capture a large fraction of a CME in the inner heliosphere within a single pointing. As a CME propagates outward, its angular extent increases, and an appropriately smaller SKA-Low sub-station beam can be selected to match the CME size while still keeping the Sun at a primary-beam null. Although sub-station observations offer reduced sensitivity, the large number of antenna elements in the SKA array ensure sufficient sensitivity for heliospheric measurements. Optimal heliospheric observations with SKA-Low will therefore require a trade-off between FoV and sensitivity, exploiting its adaptive FoV capability.

\subsection{Calibration and imaging requirements}\label{subsec:helio-calib}
Calibration of heliospheric observations will largely follow standard astronomical procedures. Although the Sun is placed at a primary-beam null, residual solar contamination may still arise due to beam chromaticity or because the null is narrow compared to the solar disk. A key calibration requirement is therefore the removal of solar contamination from standard bandpass and gain calibrators. This necessitates additional steps in the calibration strategy, similar to those implemented for MeerKAT solar observations (see Section~\ref{subsec:meerkat_calibraton}). To minimise contamination from the solar disk, only baselines longer than $200\lambda$ are used during calibration. If compact, bright sources are present on the Sun, direction-dependent calibration and source peeling are required to remove their contribution. Such source peeling must also be applied to the target field.

There is no specific tuning required for imaging parameters. But three different heliospheric techniques, IPS, FR, and angular broadening observations, required three types of imaging products -- i) IPS observations require high time-resolution (sub-second) snapshot continuum images, formed by averaging over spectral bands smaller than scintillation bandwidth, ii) FR measurements and rotation measure (RM) synthesis require spectroscopic imaging at the highest available spectral resolution with temporal integrations of approximately 15–20 minutes, and iii) imaging over timescales of tens of minutes and bandwidths of a few tens of MHz is sufficient for angular broadening measurements generated using ``uniform" weighting to maximise angular resolution. Among these, functionality for the IPS data products may need to be developed or included from outside SKA SDP (Figure \ref{fig:proposed_pipeline}), while the other two are expected to be available in the SKA SDP.

\section{Making solar and heliospheric radio observations mainstream}
\label{sec:ska_next_steps}
Solar science is inherently multi-wavelength in nature. Observational signatures of the most active solar phenomena span the waveband from X-rays to radio. It is well recognized that radio imaging observations tend to provide information complementary to that available from other wavebands \citep[e.g.][]{McLean_Labrum_1985, Gary_Keller_2004, Pick2008}. Their adoption in the larger solar community has, however, remained limited due to a combination of many reasons -- limited availability of good-quality imaging observations, 
challenging data processing and difficulties in the interpretation of the final data products. This section discusses these aspects and presents some suggestions for the way forward. 
Addressing these aspects adequately is essential for ensuring science outcomes from SKA solar and heliospheric observations.
\begin{figure}[!htbp]
    \centering
    \includegraphics[trim={0cm 7cm 0cm 0cm},clip,width=0.95\linewidth]{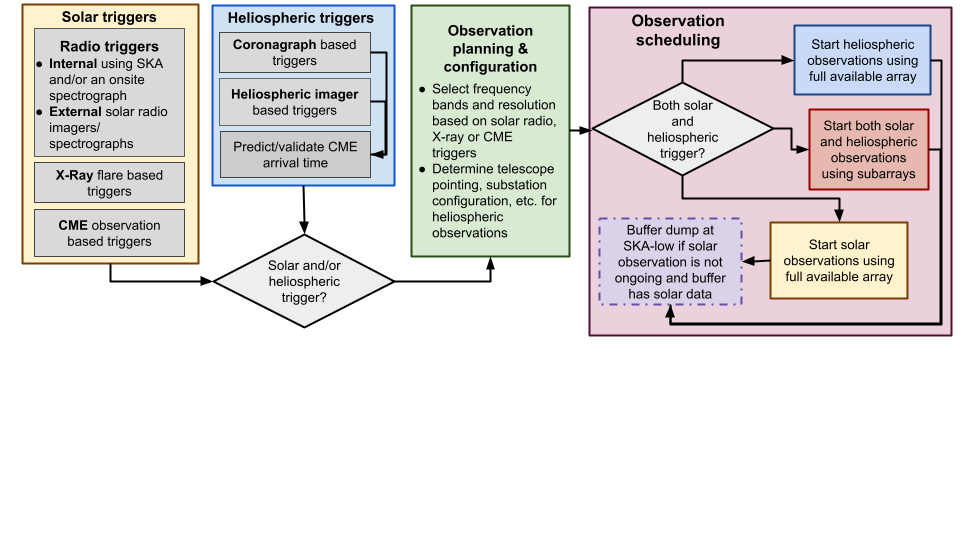}
    \caption{A proposed triggering framework for solar and heliospheric observation using SKA telescopes.}
    \label{fig:trigger}
\end{figure}

\subsection{A triggering framework for efficient observation}
The requirement for triggered solar and heliospheric observations is discussed in Sections \ref{subsec:triggered-obs} and \ref{subsec:helio-obs-strategy} respectively. We propose a unified triggering framework for both SKA telescopes that supports coordinated solar and heliospheric observations (Figure \ref{fig:trigger}). 
This system will incorporate radio (both internal and external) as well as non-radio solar triggers for initiating observations of solar radio bursts, flares, and CMEs. 
Heliospheric triggers will be derived using near–real-time coronagraph and heliospheric imager data and forecasts of CME arrival at specific heliocentric distances. A joint observation planning and configuration module will combine these triggers to determine telescope configuration, observing frequency bands, and pointing strategies.
When only the solar or heliospheric trigger is raised, the full array, or an appropriate part of it, can be configured for the desired observation.
In case both solar and heliospheric observations are simultaneously triggered, the array can be split into subarrays to meet the observing requirements.
Such a framework is essential for enabling simultaneous, high-quality solar and heliospheric observations during space-weather events and also ensures efficient use of precious SKA observing time.

\subsection{A framework for science-ready solar and heliospheric data}
\label{subsec:solar-mainstream}
\begin{figure}[!htbp]
    \centering
     \includegraphics[trim={0cm 0cm 0cm 1cm},clip,width=\linewidth]{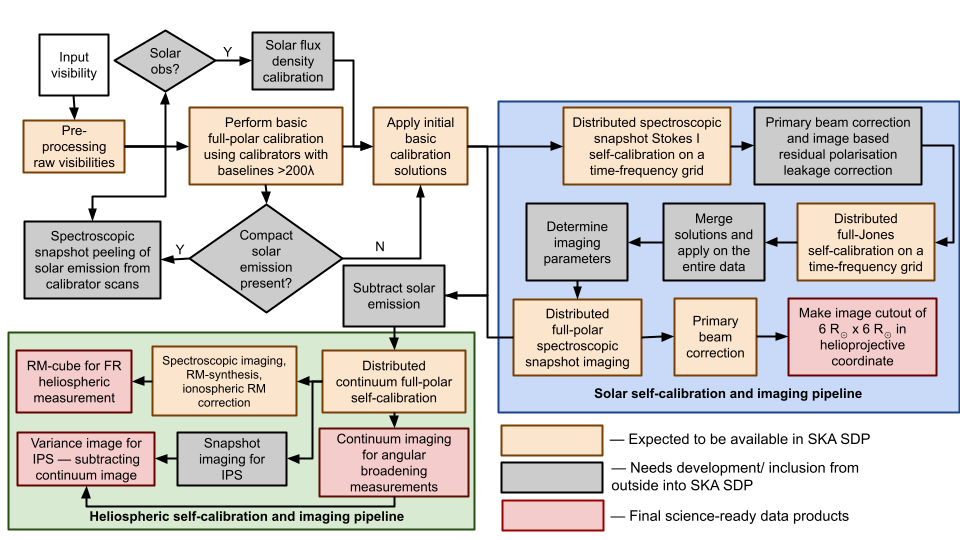}
    \caption{A proposed calibration and imaging pipeline for SKA solar and heliospheric observations. Orange boxes highlight the capability already available or expected to be available in SKA SDP, and grey boxes show the components that need to be developed for the solar pipeline.}
    \label{fig:proposed_pipeline}
\end{figure}

The use of imaging in solar radio studies has historically been comparatively limited, and the vast majority of studies rely on dynamic spectra from spectrographs. This stems from multiple factors -- until recently, the best radio interferometers were optimized for synthesis imaging and fell short of the demanding needs of solar imaging; and transforming visibilities into images required substantial effort for relatively modest scientific returns. The latter arose from the steep learning curve involved and was further hindered by the available analysis tools not being suitable for low-effort, large-scale automated processing, in stark contrast with what is available at other wavebands. 

With the advent of the SKA and its precursors and pathfinders, the instrumental limitations have been largely overcome. 
The challenge of reducing the barrier to adoption of radio observables, by making solar radio imaging as accessible and straightforward for the community as imaging of data from space-based missions, still needs to be met.
As discussed in Section \ref{sec:cal_imaging}, solar calibration and imaging are intrinsically complex. Experience gained from unsupervised calibration and imaging pipelines developed for the SKA precursors and pathfinders — such as P-AIRCARS, SIMPL, and MeerSOLAR — provides a strong foundation for the SKA solar imaging pipeline. As an illustration, Figure \ref{fig:proposed_pipeline} presents a proposed end-to-end calibration and imaging workflow to deliver science-ready data products (red boxes). Components shown in orange are expected to be available within the SKA SDP, while those in grey are likely to require additional development or external integration to support solar-specific processing. We note that the usefulness of such a pipeline will extend well beyond solar and heliospheric science. It will be directly relevant for removing solar contamination from any daytime astronomical observations using the SKA.

\subsection{Developing analysis and modelling tools}
The final step toward making solar and heliospheric radio observations truly mainstream is to provide the community with robust modelling, simulation, and interpretation tools to translate radio images in to physically meaningful parameters.
Significant progress is already being made through efforts such as computationally efficient gyrosynchrotron \citep{Kuznetsov2021} and thermal emission models \citep{Fleishman2021} and tools like the GX Simulator \citep{Nita2023}. 
Efforts towards developing tools for modelling coronal propagation effects and quantifying their impacts in terms of angular broadening and refraction, and also changes in the polarization properties of the radio emission, will also be very timely.
While developing such tools is not formally within the SKA’s remit, the SKA solar and heliospheric science working group should prioritize fostering the development of user-friendly and powerful modelling and interpretation frameworks, particularly incorporating radio observations as modelling constraints, to enable broad and effective use of SKA data by the wider solar and heliospheric community.

\section{Conclusions}
The SKA offers exceptional opportunities to advance practically all areas of astrophysics, including solar and heliospheric science, as emphasized throughout this book.
Actualizing SKA's potential for solar and heliospheric science  will require careful planning of observations, calibration, imaging, and data analysis to both capture major solar and space weather events and deliver science-ready data products and tools to the community. 
This chapter highlights the aspects specific to this science area. Many of these requirements and/or functionalities are not shared by most other science areas and hence special care needs to be taken to ensure that these needs are met by the SKA telescopes.
We present a review of the progress achieved on technical fronts of relevance to solar and heliospheric science with SKA precursors and pathfinders.
Building on this experience, where pertinent, we offer practical recommendations for the SKA telescopes and the solar and heliospheric science community to enable effective and impactful SKA solar and heliospheric observations.


\section{Acknowledgements}
This work presents observations from the MWA from Inyarrimanha Ilgari Bundara, the CSIRO Murchison Radio-astronomy Observatory. We acknowledge the Wajarri Yamaji people as the traditional owners and native title holders of the observatory site. This work also presents observations from MeerKAT radio telescope operated by the South African Radio Astronomy Observatory, which is a facility of the National Research Foundation, an agency of the Department of Science, Technology, and Innovation.  This work shows data from the International LOFAR Telescope (ILT). LOFAR is designed and constructed by ASTRON. D.O., S.D., P.M., S.M., and D.P. acknowledge support from the Department of Atomic Energy, Government of India, under project 12-R\&D-TFR-5.02-0700. D.K. acknowledges support for this research by the NASA Living with a Star Jack Eddy Postdoctoral Fellowship Program, administered by UCAR’s Cooperative Programs for the Advancement of Earth System Science (CPAESS) under award 80NSSC22M0097. J.Y acknowledges the use of data from the DART OF Chinese Meridian Project and Beijing R\&D project Z251100000725011.

\bibliographystyle{abbrvnat-maxbibnames4}
\bibliography{ref} 

\end{document}